\documentclass[showpacs,amsmath,amssymb]{revtex4}
 \usepackage{amsthm}
\usepackage{graphicx} 
\newtheorem{thm}{Theorem}[section] 
\newtheorem{lemma}[thm]{Lemma}

\newtheorem*{lem21}{Lemma II.1}
\theoremstyle{remark} 

\newcommand{\C}{\mathbb{C}}

\newcommand{\E}{{\mathcal E}}

\newcommand{\tr}{{\rm tr}}

\newcommand{\bra}[1]{\langle#1|}

\newcommand{\ket}[1]{|#1\rangle} 
\newcommand{\braket}[2]{\langle#1|#2\rangle} 
\newcommand{\ketbra}[2]{|#1\rangle \langle#2|} 
\newcommand{\vstrut}{{\rule{0in}{.14in}}}

\renewcommand{\v}{{\rm \bf v}}

\newcommand{\w}{{\rm \bf w}}

\usecounter{figure}
\newcommand{\B}{\mathcal{B}}

\renewcommand{\E}{\mathcal{E}}

\renewcommand{\L}{\mathcal{L}}
\begin{document}

\title{Overcoming a limitation of deterministic dense coding with a non-maximally entangled initial state}

\author{P.~S.~Bourdon}
\affiliation{Department of Mathematics,\\ Washington and Lee University, Lexington, VA 24450} 
\email{pbourdon@wlu.edu}
\author{E.~Gerjuoy}
\affiliation{Department of Physics and Astronomy,\\ University of Pittsburgh, Pittsburgh, PA 15260} 
\email{gerjuoy@pitt.edu}
\date{\today}

\begin{abstract} Under two-party deterministic dense-coding, Alice communicates (perfectly distinguishable) messages to Bob via a qudit from a pair of entangled qudits in pure state $\ket{\Psi}$. If $\ket{\Psi}$  represents a maximally entangled state (i.e., each of its Schmidt coefficients is $\sqrt{1/d}$)  then Alice can convey to Bob one of $d^2$ distinct messages.  If  $\ket{\Psi}$ is not maximally entangled, then Ji {\it et al.}\ [Phys.\ Rev. A\ {\bf 73}, 034307 (2006)] have shown that under the original deterministic dense-coding protocol, in which messages are encoded by  unitary operations performed on Alice's qudit,  it is impossible to encode $d^2-1$  messages.  Encoding $d^2-2$ is possible; see, e.g., the  numerical studies by  Mozes {\it et al.}\ [Phys.\ Rev. A {\bf 71}, 012311 (2005)].   Answering a question raised by Wu\ {\it et al.}\ [Phys. Rev.\ A\ {\bf 73}, 042311 (2006)],  we show that when  $\ket{\Psi}$ is not maximally entangled, the communications limit of  $d^2-2$ messages persists even when the requirement that Alice encode by unitary operations on her qudit is weakened to allow encoding by more general quantum  operators.   We then describe a dense-coding protocol that can overcome this limitation with high probability, assuming the largest Schmidt coefficient of $\ket{\Psi}$ is sufficiently close to $\sqrt{1/d}$.   In this protocol, $d^2-2$ of the messages are encoded via unitary operations on Alice's qudit and the final $(d^2-1)$-th message is encoded via a (non trace-preserving) quantum operation.\end{abstract}

\pacs{03.67.Hk,03.65.Ud,03.67.Mn}

 \maketitle

\section{Introduction}  We assume that Alice and Bob, located some distance apart, each initially controls one qudit from an entangled pair in pure state $\ket{\Psi}$.      Deterministic dense coding, originated by Bennett and Wiesner \cite{BW}, allows Alice to send to Bob, via her qudit, one of up to $d^2$ perfectly distinguishable messages.   Under the original protocol for deterministic dense coding, to send a message to Bob, Alice first applies to her qudit a local unitary operation selected from a group of  possible  ``encoding unitaries''.  She then sends her qudit via a {\it d}-dimensional noiseless quantum channel to Bob, who---with both qudits now in his possession---performs a measurement which reveals the particular encoding unitary operation that Alice had performed, i.e., reveals the particular message Alice had chosen to send.  In order that these messages be perfectly distinguishable by Bob (the hallmark of {\it deterministic} dense coding), the set of encoding unitaries that Alice may apply to her qudit must produce a set of orthogonal states in the state-space of the two-qudit system that Alice and Bob share. Since that state-space is of dimension $d^2$ she can send at most  $d^2$ messages to Bob by this process. \ She can send as many as $d^2$ messages only when $\ket{\Psi}$ is a maximally entangled state, meaning that its Schmidt coefficients are identical (each equaling $\sqrt{1/d}$).

 When $\ket{\Psi}$ is not maximally entangled,  Alice cannot send $d^2$ messages via the original dense-coding protocol; in fact, Ji {\it et al.}\ \cite{J} establish that she can send at most $d^2-2$ messages (a result suggested by numerical data in \cite{M}).    Wu {\it et al.}\ ask \cite[p.\ 10]{WCSG} whether Alice's inability to create $d^2-1$ messages (when $\ket{\Psi}$ is not maximally entangled) reflects a limitation of the type of unitary encoding employed in the original protocol.   We answer this question here, showing the limitation to $d^2-2$ messages persists even when Alice uses as encoding operations the most general quantum operations possible. Alice performs a quantum operation on her qudit via the following three-step process: (i) she pairs her qudit with an ancillary quantum particle (of dimension $\ge 2$ but otherwise arbitrary), (ii) she then applies a unitary operation to the pair, and (iii) she then either measures the ancilla or chooses not to measure the ancilla.  Mathematically speaking, if in step (iii), she chooses not to measure the ancilla, then the quantum operation Alice has performed is trace preserving.  If in step (iii), she does measure the ancilla, then her quantum operation may not preserve trace.  For further information about quantum operators the reader may consult \cite[Chapter 8]{NC}.
 
  This paper is organized as follows. In the next section, we describe the mathematical framework for our results, providing a  discussion of deterministic dense coding with quantum encoding operators.   In Section 3, we show that  when the initial state $\ket{\Psi}$ of their two-qudit system is not maximally entangled, Alice cannot send $d^2-1$ messages to Bob under deterministic dense coding even when she may use quantum operators to encode messages.   However, in Section 4, we introduce a new dense-coding procedure that allows some non-maximally entangled states $\ket{\Psi}$ to support, with high probability, communication of $d^2-1$ perfectly distinguishable messages, with $d^2-2$ of the messages encoded by unitary operations on Alice's qudit and the final $(d^2-1)$-th message encoded via a (non trace-preserving) quantum operation.  In fact, for any probability $p$ as close as  desired to  $1$, we show there is a non-maximally entangled state $\ket{\Psi}$ that will support, with probability exceeding $p$, communication of $d^2-1$ messages under our protocol.  Our protocol is designed so that Bob will never misinterpret a message; rather, there is a small chance he will receive no message.  He will receive no message only when Alice wishes to send the $(d^2-1)$-th message and her encoding procedure fails.     We conclude the paper with a  detailed example illustrating how Alice and Bob can, with probability exceeding 97\%, use a two-qubit system in state $\ket{\Psi} =  \frac{9}{4\sqrt{10}} \ket{00} + \frac{\sqrt{79}}{4\sqrt{10}}\ket{11}$ as a resource for the communication of three perfectly distinguishable messages (via a two-dimensional noiseless quantum channel). In this example, the probability of success rises to over 98\% if Alice and Bob are willing to tolerate a small chance of Bob's incorrectly interpreting a message.
  
  We note that the $d=2$ case of the result of Section 3 of this paper appears in Appendix A of \cite{CB}.

\section{Background}

\subsection{The initial state} Let $H = H_A\otimes H_B$ be the state space of the two-qudit system that Alice and Bob share, where  $(\ket{0}_A, \ket{1}_A, \ldots, \ket{d-1}_A)$ is an orthonormal basis for  $H_A$ and  $(\ket{0}_B, \ket{1}_B, \ldots, \ket{d-1}_B)$ is an orthonormal basis for  $H_B$. Note $H$ has orthonormal basis $\B = \{\ket{ij}: 0 \leq i, j \leq d-1\}$, where we have used  $\ket{ij}$ as a convenient substitute for $\ket{i}_A\ket{j}_B$.  We assume that  the initial state $\ket{\Psi}$  of  Alice and Bob's  two-qudit system has Schmidt representation 
\begin{equation}\label{GQDS}
\ket{\Psi} = \sum_{j=0}^{d-1} \sqrt{\lambda_j} \ket{jj},
\end{equation}
where the Schmidt coefficients $\sqrt{\lambda_0}$, $\sqrt{\lambda_1}$, \ldots, $\sqrt{\lambda_{d-1}}$ have squares summing to $1$ (which assures normalization),  and where we assume, without loss of generality, that 
 \begin{equation}\label{orderasmpt}
 \lambda_0 \ge \lambda_1\ge  \ldots \ge \lambda_{d-1}\ge
0.
 \end{equation}
 We will frequently describe the initial state in terms of its density operator $\ketbra{\Psi}{\Psi}$. 
 
 Wu {\it et al.}\ \cite[Section IV.B]{WCSG} establish that when $\ket{\Psi}$ of (\ref{GQDS}) allows Alice's sending to Bob $L$ perfectly distinguishable messages (using any encoding scheme), then
\begin{equation}\label{WCSGB}
 \lambda_0 \le \frac{d}{L}.
\end{equation}
Suppose, e.g., that $L > d(d-1)$, then the preceding inequality yields $\lambda_0 < 1/(d-1)$.  Since $\sqrt{\lambda_0}$ is the largest Schmidt coefficient of $\ket{\Psi}$ and $\sum_{j=0}^{d-1} \lambda_j = 1$, it follows that if $\lambda_0 < 1/(d-1)$, then every Schmidt coefficient in (\ref{GQDS}) is nonzero.  In particular, if we assume Alice is able to deterministically send $L= d^2-1$ messages,  then, because $d^2-1 > d(d-1)$, all Schmidt coefficients of $\ket{\Psi}$ must be nonzero. {\it For the remainder of this paper, we assume all Schmidt coefficients of $\ket{\Psi}$ are nonzero.}

\subsection{Encoding operations}  

For any vector space $W$, we let $\L(W)$ denote the vector space of all linear operators on $W$.

 Recall that Alice encodes messages for Bob, physically speaking, by applying a unitary operation to either (i) her qudit or, more generally, (ii) to her qudit paired with an ancillary particle, perhaps measuring the ancilla afterwards.    In either case, Alice's encoding action may be represented mathematically  by a quantum operator  applied to $\ketbra{\Psi}{\Psi}\in \L(H)$: 
\begin{equation}\label{QO}
\ketbra{\Psi}{\Psi} \mapsto \sum_{j=0}^{N-1} (K^{(j)}\otimes I_B) \ketbra{\Psi}{\Psi}(K^{(j)} \otimes I_B)^\dagger,
\end{equation} 
   where $N$ is a positive integer and the $K^{(j)}$'s are {\it Kraus operators} in $\L(H_A)$ satisfying
 \begin{equation}\label{NNTP}
    \sum_{j=0}^{N-1} (K^{(j)})^\dagger K^{(j)} \le I_A.
\end{equation}
In case (i), the sum on the right of  (\ref{QO}) has only one summand and $K^{(0)} = U$, where $U$ is a unitary operator on $H_A$.  Thus, in case (i) the inequality (\ref{NNTP}) is an equality. It's also an equality  in case (ii)  (see, e.g., Appendix B) provided Alice does not measure the ancilla.  

As we explain in Section~\ref{QODC} below, it's easy to see that Alice's ability to measure the ancilla  can never be used to increase the number of messages she can send to Bob through  deterministic dense coding.   Thus for now, we will assume that Alice does not measure the ancillary particle, which means that in either case (i) or case (ii), Alice's encoding action is described by a quantum operator $\E$ having the {\it operator sum representation} 
\begin{equation}\label{QOP}
\E(\rho) =   \sum_{j=0}^{N-1} (K^{(j)}\otimes I_B) \rho(K^{(j)} \otimes I_B)^\dagger, \quad \rho\in L(H),
\end{equation}
where the $K^{(j)}$'s satisfy
\begin{equation}\label{TPC}
    \sum_{j=0}^{N-1} (K^{(j)})^\dagger K^{(j)} = I_A,
\end{equation}
making  $\E$ {\it trace preserving}. 

  The quantum operator $\E$ of (\ref{QOP}) has many different operator sum representations (see, e.g., \cite[Theorem 8.2]{NC}).  Among these representations, there is one for which the number $N$ of Kraus-operator elements assumes its minimum possible value $m$.  This number $m$ is the {\it Kraus rank} of $\E$ and it is easy to see that any group of exactly $m$ Kraus-operator elements  representing $\E$ must be linearly independent in $\L(H_A)$    (which is equivalent to the linear independence of  $K^{(0)}\otimes I_B$, $K^{(1)}\otimes I_B$, \ldots, $K^{(m-1)}\otimes I_B$ in  $\L(H)$).    
  
  We now state an important Lemma for our work; its (short) proof occupies Appendix A.
  
  \begin{lemma}\label{LIL}  Suppose that all the Schmidt coefficients of $\ket{\Psi}$ in the representation (\ref{GQDS}) are nonzero and that  $K^{(0)}, K^{(1)}, \ldots, K^{(m-1)}$ are linearly independent in $\L(H_A)$; then $(K^{(0)}\otimes I_B)\ket{\Psi}, (K^{(1)}\otimes I_B)\ket{\Psi}, \ldots, (K^{(m-1)}\otimes I_B)\ket{\Psi}$ are linearly independent vectors in $H$.
  \end{lemma}
    
  \subsection{Perfect distinguishability}

   In order to send $L$ perfectly distinguishable messages to Bob, Alice must be able to perform $L$ encoding operations on her qudit (perhaps paired with an ancilla) with each such operation producing a message that Bob will recognize and never mistake for a message corresponding to another encoding operation.    Each  of the $L$ encoding operations results in a density-operator description of the two-qudit system that Alice and Bob share: $\rho_0, \rho_1, \ldots, \rho_{L-1}$. The perfectly distinguishability of the corresponding messages means that 
\begin{equation}\label{PDM}
   \rho_i\rho_j = 0 \quad {\rm whenever}  \quad i\ne j;
   \end{equation}
i.e., the supports of the density operators representing messages must be orthogonal.  (For a formal proof that orthogonality of the supports provides perfect distinguishability, see \cite[Theorem 1]{Fg}.)

  Observe that the support of the density operator $\E(\ketbra{\Psi}{\Psi})$ representing the message encoded by the trace-preserving quantum operator (\ref{QOP}) is precisely the linear span of
  \begin{equation}\label{SS}
\{(K^{(0)} \otimes I_B)\ket{\Psi}, (K^{(1)} \otimes I_B)\ket{\Psi}, \ldots, (K^{(N-1)} \otimes I_B)\ket{\Psi}\}.
\end{equation}
Thus the perfect distinguishability of the messages Alice produces via her quantum encoding operations amounts  to the following at the Kraus-operator level:  messages produced by distinct quantum operations $\E_1$ and $\E_2$ are perfectly distinguishable if and only if whenever $K_1$ is a Kraus operator for $\E_1$  and  $K_2$ is a Kraus operator for $\E_2$, then $\bra{\Psi}(K_1\otimes I_B)^\dagger(K_2\otimes I_B)\ket{\Psi}= 0$.

\subsection{Decoding messages}

  Suppose that Alice is able to encode $L$ perfectly distinguishable messages, represented by density operators $\rho_0, \rho_1, \ldots \rho_{L-1}$.    For each $j= 0, 1, \ldots, L-1$, let $S_j$ be the support of $\rho_j$, so that by (\ref{PDM}), $\{S_j\}_{j=0}^{L-1}$ is a collection of pairwise orthogonal subspaces of  $H$.  To send message $j$ to Bob, Alice performs the quantum operation $\E$ that creates the state  $\rho_j$ and sends her qudit to Bob through a noiseless channel, keeping any ancillary particle she may have used in executing $\E$.  To decode Alice's message,  Bob simply performs a projective measurement described by the observable
$$
\sum_{j=0}^{L-1} j P_{S_j},
$$
which is equivalent to the observable $\sum_{j=0}^{L-1} j (P_{S_j}\otimes I_a)$, where $I_a$ is the identity on the Hilbert space $H_a$ of the ancilla.
The pairwise orthogonality of  the subspaces $S_j$  ensures Bob will measure $j$  precisely when message $j$ has been sent, i.e.,  precisely when Alice has created the ``message state'' $\rho_j$.

\subsection{Ancilla measurement}\label{QODC}

     Suppose that Alice pairs her qudit $A$ with an ancillary $N$-level particle $a$.   The Hilbert space for the $Aa$ pair is $H_A\otimes H_a$, with orthonormal basis $\{\ket{i}_A\ket{j}_a: i = 0, 1, 2, \ldots, d-1; j=0, 1, 2, \ldots, N-1\}$.   Assume, as above, that Alice's particle $A$ is entangled with Bob's particle $B$ and their two-qudit system is in state $\ket{\Psi}$ given by (\ref{GQDS}).   Assume that $a$ is in state $\ket{0}_a$.   Suppose Alice performs a unitary operation $U$ on the pair $Aa$.     The effect of Alice's unitary operator $U$ on the state $\ket{\Psi}\ket{0}_a$ may be described as follows (see Appendix B):
\begin{equation}\label{ROU}
\ket{\Psi}\ket{0}_a\mapsto  \sum_{j=0}^{N-1} \left(  K^{(j)} \otimes I_B \right)\ket{\Psi}\otimes \ket{j}_a,\end{equation}
where the $K^{(j)}$'s are Kraus operators that satisfy (\ref{TPC}).

Forming the density operator corresponding to (\ref{ROU}) and taking the partial trace over the ancillary system produces the density-operator on the right of  (\ref{QO}), which describes the message state that Alice creates using $U$.  Denote by $\E$ the corresponding trace-preserving quantum operation on $\L(H)$---it has the form (\ref{QOP}) and  the density operator $\E(\ketbra{\Psi}{\Psi})$ for the encoded message has support equal to the linear span of the set (\ref{SS}).    Suppose that Alice applies $U$ and then performs a measurement of the ancilla $a$ described by the collection $\{M_x\}$ of measurement operators.   Assuming $y$ is the outcome of the measurement and recalling that the application of $U$ to the state  $\ket{\Psi}\ket{0}_a$  yields the state (\ref{ROU}), we see the state of the  $ABa$ system after measurement is 
\begin{equation}\label{SAM}
c\sum_{j=0}^{N-1} \left(  K^{(j)} \otimes I_B \right)\ket{\Psi}\otimes M_y\ket{j}_a,
\end{equation}
where $c$ is a normalizing constant.   Since $M_y\ket{j}_a = \sum_{i=0}^{N-1} \alpha_{ij}^{(y)} \ket{i}_a$  for some collection of scalars $\alpha_{ij}^{(y)}$,  expression (\ref{SAM}) may be written 
$$
c\sum_{i=0}^{N-1} \left(\sum_{j=0}^{N-1}\alpha_{ij}^{(y)}  \left(K^{(j)} \otimes I_B \right)\ket{\Psi}\right)\otimes\ket{i}_a,
$$
which corresponds to a density operator  $\rho$ on $\L(H)$ whose support will be contained in that of $\E(\ketbra{\Psi}{\Psi})$, because the support of $\rho$ consists of linear combinations of vectors of the form  $\sum_{j=0}^{N-1}\alpha_{ij}^{(y)}  \left(K^{(j)} \otimes I_B \right)\ket{\Psi}$, each one of which is in the support of $ \E(\ketbra{\Psi}{\Psi})$.     Hence if Alice measures the ancilla before she sends her message to Bob , then he will still receive the intended message.  Equally important is that because the measurement operators $\{M_x\}$ satisfy the completeness relation $\sum_x M_x^\dagger M_x = I_a$, Alice cannot predetermine some proper subspace $S$ of the support of $\E(\ketbra{\Psi}{\Psi})$ and use measurement of the ancilla to  produce {\em with certainty} a state of the $AB$ system whose density-operator description $\rho$ has support contained in $S$.   It follows that Alice cannot use measurement of ancillary particles during the encoding process to increase the number of messages she may send to Bob via deterministic dense coding. 

We now turn to our main results. 

\section{Alice Cannot encode $d^2-1$ messages when $\ket{\Psi}$ is not maximally entangled}\label{CBJ}

We suppose that Alice can use  quantum operators   $\E_0, \E_1, \ldots, \E_{d^2-2}$ to encode $d^2-1$ perfectly distinguishable messages for Bob and prove that their initial two-qudit state $\ket{\Psi}$ must be maximally entangled.  We have established that we may assume, without loss of generality, that Alice's encoding quantum operators are trace-preserving.    For $j = 0, 1, \dots, d^2-2$, let $m_j$ be the Kraus rank of $\E_j$.  Thanks to Lemma~\ref{LIL} and the discussion following Eq.\ (\ref{WCSGB}), the support of the density operator $\E_j(\ketbra{\Psi}{\Psi})$ must have dimension $m_j$.   We know that perfect distinguishability of messages means that the supports of the density operators   $\E_0(\ketbra{\Psi}{\Psi})$, $\E_1(\ketbra{\Psi}{\Psi})$, \ldots, $\E_{d^2-2}(\ketbra{\Psi}{\Psi})$ must be pairwise orthogonal.  These supports are subspaces of the $d^2$ dimensional space $H = H_A\otimes H_B$.  Thus 
\begin{equation}\label{ENSL}
  \sum_{j=0}^{d^2-2} m_j \le d^2.
 \end{equation}

  Since $m_j\ge 1$ for each $j$, the preceding inequality shows that $m_j =1$ for all but $1$ of the $j$'s and for the remaining $j$ value, either $m_j =1$ or $m_j = 2$.   Note that if $m_j=1$, then since $\E_j$ is trace preserving (i.e., its Kraus-operator elements satisfy (\ref{TPC})), we see that $\E_j$  is an original-protocol unitary encoding operation.   Thus if  $m_j = 1$ for  every  $j$, then Alice can send to Bob $d^2-1$ messages via original-protocol unitary encoding and, as we indicated earlier, Ji {\it et al.}\ have shown that in this case $\ket{\Psi}$ must be maximally entangled.   Thus,  to complete the argument, we must show that $\ket{\Psi}$ must also be maximally entangled in case $m_j = 2$ for some $j$ and the rest of the $m_j$'s equal $1$.   
  
  Without loss of generality, we assume that $m_{d^2-2} = 2$ so that $\E_{d^2-2}$ may be expressed in the form
  $$
  \E_{d^2-2}(\rho) =  (K^{(0)}\otimes I_B)\rho(K^{(0)} \otimes I_B)^\dagger  + (K^{(1)}\otimes I_B)\rho(K^{(1)} \otimes I_B)^\dagger,
  $$
  where $K^{(0)}$ and $K^{(1)}$ are linearly independent Kraus operators satisfying
\begin{equation}\label{tp}
(K^{(0)})^\dagger K^{(0)} + (K^{(1)})^\dagger K^{(1)} = I_A.
\end{equation}
Each of the remaining encoding operators $\E_0, \E_1, \ldots, \E_{d^2-3}$ is an original protocol unitary operation:
$$
\E_j(\rho) = (U^{(j)}\otimes I_B)\rho   (U^{(j)}\otimes I_B)^\dagger,
$$ 
where for $j = 0, 1, \ldots, d^2-3$, $U^{(j)}$ is a unitary operator on $H_A$.     Perfect distinguishability  of the messages produced by $\E_0, \ldots, \E_{d^2-3}$, i.e., pairwise orthogonality of support sets of the corresponding density operators, means that  
\begin{equation}\label{UMS}
\{ (U^{(0)} \otimes I_B)\ket{\Psi}, (U^{(1)} \otimes I_B)\ket{\Psi}, \ldots, (U^{(d^2-3)} \otimes I_B)\ket{\Psi}\}
\end{equation}
 is an orthogonal set in $H$.  Moreover, the support of $\E_{d^2-2}(\ketbra{\Psi}{\Psi})$, representing the ``final message'', must be orthogonal to the subspace of $H$ spanned by the vectors in the set (\ref{UMS}). Equivalently, each of 
\begin{equation}\label{KST}
 \ket{\phi_0} \equiv (K^{(0)}\otimes I_B)\ket{\Psi} \quad {\rm and} \quad   \ket{\phi_1}\equiv (K^{(1)}\otimes I_B)\ket{\Psi}
 \end{equation}
  is orthogonal to every vector in the set (\ref{UMS}).   

In fact, we can even assume the two  vectors $ \ket{\phi_0}$ and $\ket{\phi_1}$ representing the  $(d^2-1)$-th message
 are orthogonal to each other. We justify this claim in Appendix C.  Thus, henceforth we assume that the two Kraus states $\ket{\phi_0}$ and $\ket{\phi_1}$   representing the  the $(d^2-1)$-th message  are orthogonal to each other as well as to each element of the set (\ref{UMS}).  Note well that $\ket{\phi_0}$ and $\ket{\phi_1}$ are linearly independent vectors (by Lemma~\ref{LIL}); thus, in particular, neither is the zero vector.
 
 Though they are orthogonal,   $\ket{\phi_0}$ and $\ket{\phi_1}$  aren't normalized. Since $\braket{\phi_0}{\phi_0} + \braket{\phi_1}{\phi_1} = 1$ (via (\ref{tp})), if we set
\begin{equation}\label{xdef}
x = \braket{\phi_0}{\phi_0},
\end{equation}
then $0 < x < 1$ (both $\ket{\phi_0}$ and $\ket{\phi_1}$ are nonzero) and the pair of ``Kraus states'' spanning the support of  $\E_{d^2-2}(\ketbra{\Psi}{\Psi})$ are
$$
\ket{\phi_0}/\sqrt{x} \quad \text{and} \quad  \ket{\phi_1}/\sqrt{1-x}.
$$
We assume, without loss of generality,  that $0 < x\le 1/2$  (otherwise we can just switch labels on $\ket{\phi_0}$ and $\ket{\phi_1}$. 

We complete the proof by establishing that the existence of the following orthonormal subset of $H$ forces all Schmidt coefficients of $\ket{\Psi}$ to have the same value, so that $\ket{\Psi}$ is maximally entangled:
\begin{equation}\label{KMS}
\{(U^{(0)} \otimes I_B)\ket{\Psi},  (U^{(1)} \otimes I_B)\ket{\Psi},  \ldots, (U^{(d^2-3)} \otimes I_B)\ket{\Psi},   \ket{\phi_0}/\sqrt{x},\ \ket{\phi_1}/\sqrt{1-x}\}.
\end{equation}

We view the operators $K^{(q)}$  of (\ref{KST}) as well as the operators $U^{(n)}$ of (\ref{UMS})---all of which are operators on the $d$-dimensional Hilbert space $H_A$---as $d\times d$ matrices with respect to the basis $(\ket{0}_A, \ket{1}_A, \ldots, \ket{d-1}_A)$, and we let $u^{(n)}_{ij}$ and $k^{(q)}_{ij}$ denote the entries of these matrices.    For $j = 0, 1, 2, \ldots, d-1$ and $q = 0, 1$,  let 
$$
_jK^{(q)}
$$ 
denote the $j$-th column of the $d\times d$ matrix $K^{(q)}$. Note well that the initial column of $K^{(q)}$ is $_0K^{(q)}$.   From Eq.\  (\ref{tp}), we conclude
\begin{equation}\label{ftp}
\braket{_iK^{(0)}}{_jK^{(0)}} + \braket{_iK^{(1)}}{_jK^{(1)}} = \delta_{ij}.
\end{equation}

Define $d\times d$ matrices
\begin{equation}\label{NW}
E = \frac{1}{\sqrt{x}} K^{(0)}  \quad {\rm and}\quad W = \frac{1}{\sqrt{1-x}} K^{(1)}.
\end{equation}

We now order the basis for the Hilbert space $H_A\otimes H_B$ of the Alice-Bob system:
\begin{equation}\label{Bwo}
\B = (\ket{00}, \ket{10}, \ket{2, 0}, \ldots, \ket{d-1, 0}, \ket{01}, \ket{11}, \ldots, \ket{d-1,1}, \ldots, \ket{d-1,d-1}).
\end{equation}
Thus the basis elements are listed in $d$ groups of $d$ elements with the ordering of the groups determined by the second of the pair $\ket{ij}$ and the ordering within the groups determined by the first of the pair.  This is the ordering used by Gerjuoy {\it et al.}\ in  \cite{GWB} to form an {\it augmented message matrix} for an original-protocol  unitary encoding of messages. Gerjuoy {\it et al.}\ use the augmented message matrix to, e.g., present an alternate proof of the result of  Ji {\it et al.}\ establishing that $d^2-1$ messages cannot be produced by original-protocol unitary encoding.  We use a similar matrix $M$ below.

View $\ket{\phi_0}/\sqrt{x}$ and $\ket{\phi_1}/\sqrt{1-x}$ as column vectors---consisting, respectively, of the coordinates of $\ket{\phi_0}/\sqrt{x}$ and $\ket{\phi_1}/\sqrt{1-x}$ with respect to the basis $\B$:
$$
\left[\frac{\ket{\phi_0}}{\sqrt{x}}\right]_{ij} = \frac{\sqrt{\lambda_j}}{\sqrt{x}}k_{ij}^{(0)}\quad {\rm and} \quad  \left[\frac{\ket{\phi_1}}{\sqrt{1-x}}\right]_{ij} = \frac{\sqrt{\lambda_j}}{\sqrt{1-x}}k_{ij}^{(1)}.
$$
 The natural ordering for the entries of these vectors is provided by the ordering of the basis pairings in $\B$.  Thus,  $00$ is the initial entry (followed by $10$; $20$; \ldots; $d-1,0$), the $d+1$ entry is $01$, and the final entry is $d-1, d-1$.   Thus the first $d$ entries of $\ket{\phi_0}/\sqrt{x}$ constitute the column $(\sqrt{\lambda_0}/\sqrt{x}) (_0K^{(0)})$, the next $d$ constitute $(\sqrt{\lambda_1}/\sqrt{x}) (_1K^{(0)})$, etc.  Similarly, the first $d$ entries of $\ket{\phi_1}/\sqrt{1-x}$ constitute $(\sqrt{\lambda_0}/\sqrt{1-x})(_0K^{(1)})$, the next $d$ constitute $(\sqrt{\lambda_1}/\sqrt{1-x})( _1K^{(1)})$, etc.

 Form the $d^2\times d^2$  matrix $M$ whose first $d^2-2$ columns are, in order, the coordinates with respect to $\B$ of $(U^{(n)}\otimes I_B) \ket{\Psi}$, $n=0, 1, \ldots, d^2-3$, and whose final two columns are  $\ket{\phi_0}/\sqrt{x}$ (penultimate) and $\ket{\phi_1}/\sqrt{1-x}$.  Thus $M$ is the natural $d^2\times d^2$ matrix corresponding to the orthonormal set (\ref{KMS}).   Because the columns of $M$ constitute an orthonormal basis of $\C^{d^2}$, $M$ is unitary.

  Using the structure of $M$, we now prove that if $0\le i, j\le d-1$ and $i\ne j$, then
\begin{equation}\label{AO}
\frac{\sqrt{\lambda_i\lambda_j}}{x} \braket{_iK^{(0)}}{_jK^{(0)}} + \frac{\sqrt{\lambda_i\lambda_j}}{1-x}\braket{_iK^{(1)}}{_jK^{(1)}}  = 0.
\end{equation}
  Because the matrix $M$ is unitary, the inner product of each pair of distinct rows of $M$ is $0$. Thus if $i,j$ are distinct elements of $\{0, 1, \ldots, d-1\}$,  then upon taking the inner products of rows labeled by $si$ and $sj$ and next  summing over $s$, we have
\begin{eqnarray*}
 0 & = & \sum_{s=0}^{d-1} \left(\sqrt{\lambda_i\lambda_j} \sum_{n=0}^{d^2-3} (u^{(n)}_{si})^*u^{(n)}_{sj}\right) + \sum_{s=0}^{d-1}\frac{\sqrt{\lambda_i\lambda_j}}{x}  (k_{si}^{(0)})^*k_{sj}^{(0)} + \sum_{s=0}^{d-1}\frac{\sqrt{\lambda_i\lambda_j}}{1-x}  (k_{si}^{(1)})^*k_{sj}^{(1)}\\
  & = &\sum_{n=0}^{d^2-3} \left(\sqrt{\lambda_i\lambda_j} \sum_{s=0}^{d-1} (u^{(n)}_{si})^*u^{(n)}_{sj}\right) + \frac{\sqrt{\lambda_i\lambda_j}}{x} \braket{_iK^{(0)}}{_jK^{(0)}} + \frac{\sqrt{\lambda_i\lambda_j}}{1-x}\braket{_iK^{(1)}}{_jK^{(1)}} \\
  & = & \frac{\sqrt{\lambda_i\lambda_j}}{x} \braket{_iK^{(0)}}{_jK^{(0)}} + \frac{\sqrt{\lambda_i\lambda_j}}{1-x}\braket{_iK^{(1)}}{_jK^{(1)}} ,
  \end{eqnarray*}
  where the final equality holds because for each $n\in \{0, 1, 2, \ldots, d^2-3\}$, the matrix $U^{(n)}$ is unitary (in particular, its columns are orthogonal).  We have proved (\ref{AO}).  Continuing to assume $i\ne j$, we now combine (\ref{AO}) and (\ref{ftp}) and record the result in matrix-equation form:
\begin{equation}\label{ME}
  \begin{bmatrix}1& 1\\ \frac{\sqrt{\lambda_i\lambda_j}}{x} & \frac{\sqrt{\lambda_i\lambda_j}}{1-x}\end{bmatrix} \begin{bmatrix} \braket{_iK^{(0)}}{_jK^{(0)}}  \\ \braket{_iK^{(1)}}{_jK^{(1)}} \end{bmatrix}= \begin{bmatrix} 0\\ 0\end{bmatrix}.
\end{equation}
 The determinant of the matrix of coefficients on the left of the preceding equation is $\sqrt{\lambda_i\lambda_j}\left(\frac{1}{1-x} - \frac{1}{x}\right)$.  There are two possibilities: (i) either this determinant is $0$, in which case $x = 1/2$ (since $\lambda_i\lambda_j$ is nonzero), or (ii) this determinant is nonzero, in which case (\ref{ME}) shows that we must have 
\begin{equation}\label{CO}
  \braket{_iK^{(0)}}{_jK^{(0)}} = 0\quad {\rm and} \quad \braket{_iK^{(1)}}{_jK^{(1)}} = 0  
\end{equation}
  for all pairs of distinct $i$ and $j$ in $\{0, 1, 2, \ldots, d-1\}$.   We show that in both case (i) and case (ii), $\ket{\Psi}$ must be a maximally entangled state, completing the argument.
\subsection{  Case (i): $x = 1/2$.}

Unitarity of $M$ implies each of its rows has length one; thus, for every $i$ and $j$ in $\{0, 1,2, \ldots, d-1\}$, we have $ \sum_{n=0}^{d^{2}-1}|M_{ij,n}|^{2} =1$; equivalently,
\begin{equation}\label{Mrow}
\lambda _{j}\left[
\sum_{n=0}^{d^{2}-3}|u_{ij}^{(n)}|^{2}+\frac{|k_{ij}^{(0)}|^{2}}{x}+\frac{|k_{ij}^{(1)}|^{2}}{1-x}\right] =1.
\end{equation}
 Restricting attention to diagonal entries, we see that Eq. (\ref{tp}) reveals 
   $$
   \sum_{i=0}^{d-1}\left(|k_{ij}^{(0)}|^2 + |k_{ij}^{(1)}|^2\right) = 1
   $$
   for each $j\in \{0, 1, \ldots, d-1\}$.
   Thus, if we set 
   \begin{equation}\label{bnote}
   b_j =  \sum_{i=0}^{d-1} |k_{ij}^{(0)}|^2,\ \text{then} \ \sum_{i=0}^{d-1}|k_{ij}^{(1)}|^2 = 1-b_j.
   \end{equation}  
   
  Now sum both sides of Eq.\ (\ref{Mrow}) from $i = 0$ to $d-1$ and use the unitarity of $U^{(n)}$ for each $n$ to see that
  \begin{equation}\label{EKE}
\lambda _{j}\left[d^{2}-2+\frac{b_{j}}{x}+\frac{1-b_{j}}{1-x}\right] = d
\end{equation}
for each $j$.
 Because we are assuming $x = 1/2$ for Case (i), the expression in square brackets on the left of the preceding equation simplifies to $d^2$ and the equation yields  $ \lambda_j d^2 = d$ for every $j$, that is, $\lambda_j = 1/d$ for every $j$ and we have shown in this case that $\ket{\Psi}$ is maximally entangled.

 \subsection{ Case (ii): $0 < x < 1/2$}
 
    Equation (\ref{CO}) holds in this case so that distinct columns of the ``Kraus matrix '' $K^{(0)}$ are orthogonal, and the same is true of $K^{(1)}$.    Thus recalling the definitions of the matrices $E$ and $W$ from (\ref{NW}) as well as  the notation introduced in (\ref{bnote}), we have  
  \begin{equation}\label{NDN}
  E^\dagger E = \begin{bmatrix}\frac{b_0}{x} & 0 & 0 & \cdots & 0\\ 0& \frac{b_1}{x} & 0 & \cdots & 0\\ \vdots & \vdots & \vdots & \vdots & \vdots \\ 0 & 0 & \cdots & 0 & \frac{b_{d-1}}{x}\end{bmatrix}
\end{equation}
and
\begin{equation}\label{WDW}
  W^\dagger W = \begin{bmatrix}\frac{1-b_0}{1-x} & 0 & 0 & \cdots & 0\\ 0& \frac{1-b_1}{1-x} & 0 & \cdots & 0\\ \vdots & \vdots & \vdots & \vdots & \vdots \\ 0 & 0 & \cdots & 0 & \frac{1-b_{d-1}}{1-x}\end{bmatrix}.
\end{equation}

  We need to exploit further the structure of  the unitary  matrix $M$.  Note that Eq, (\ref{Mrow}) may be rewritten
    \begin{equation}\label{REE}
   \sum_{n=0}^{d^2-3} |u_{ij}^{(n)}|^2 + |k_{ij}^{(0)}|^2/x + |k_{ij}^{(1)}|^2/(1-x)  = \frac{1}{\lambda_j}.
   \end{equation}
Fix $i$ in Eq.\ (\ref{REE}) and sum both sides from  $j= 0$ to $j= d-1$; use the fact that the rows of each $U^{(n)}$ matrix all have length one to obtain
\begin{equation}\label{srl}
 \frac{1}{x}\sum_{j=0}^{d-1} |k_{ij}^{(0)}|^2 + \frac{1}{1-x}\sum_{j=0}^{d-1} |k_{ij}^{(1)}|^2= \sum_{j=0}^{d-1}\frac{1}{\lambda_j} - (d^2-2).
\end{equation}
Note well that it follows from the preceding equation that the diagonal entries of the matrix
\begin{equation}\label{NNDPWWD}
EE^\dagger + WW^\dagger 
\end{equation}
all have common value $\sum_{j=0}^{d-1}\frac{1}{\lambda_j} - (d^2-2)$.
We claim that every off-diagonal entry of $EE^\dagger + WW^\dagger $  is $0$.

Let $p$ and $q$ be distinct integers in $\{0, 1, 2, \ldots\ d-1\}$.  Fix $j\in\{0, 1, 2\, \ldots, d-1\}$ and take the inner product of rows $q,j$ and $p,j$ of $M$ to obtain
$$
0 = \lambda_j\left[ \sum_{n=0}^{d^2-3} u_{pj}^{(n)}(u_{qj}^{(n)})^* + \frac{1}{x} k_{pj}^{(0)} (k_{qj}^{(0)})^* + \frac{1}{1-x} k_{pj}^{(1)}(k_{qj}^{(1)})^*\right]
$$
so that
$$
0 =  \sum_{n=0}^{d^2-3} u_{pj}^{(n)}(u_{qj}^{(n)})^* + \frac{1}{x} k_{pj}^{(0)} (k_{qj}^{(0)})^* + \frac{1}{1-x} k_{pj}^{(1)}(k_{qj}^{(1)})^*.
$$
Now sum both sides of the preceding equation from $j = 0$ to $j=d-1$ and use the orthogonality of row $p$ of $U^{(n)}$ and row $q$ of $U^{(n)}$ for each $n$ to obtain
\begin{eqnarray*}
0 & =& \sum_{j=0}^{d-1}\left( \frac{1}{x} k_{pj}^{(0)} (k_{qj}^{(0)})^* + \frac{1}{1-x} k_{pj}^{(1)}(k_{qj}^{(1)})^*\right) \\
    & = & (EE^\dagger + WW^\dagger)_{pq},
   \end{eqnarray*}
   and it follows that $EE^\dagger + WW^\dagger$ is a diagonal matrix.  Using our earlier observation that all the diagonal entries of $EE^\dagger + WW^\dagger $  have common value $\gamma\equiv\sum_{j=0}^{d-1}\frac{1}{\lambda_j} - (d^2-2)$ we see
\begin{equation}\label{RowKey}
   EE^\dagger + WW^\dagger =   \gamma I.
\end{equation}

The conclusion of our argument relies upon the following observation arising from the {\it polar decomposition} (see, e.g., Theorem 2.3 on page 78 of \cite{NC}).  For any $n\times n$ matrix $Y$, we know there is a unitary matrix $U$ such that
$$
Y = U \sqrt{Y^\dagger Y} = \sqrt{YY^\dagger} U.
$$
Thus  
$$
\sqrt{YY^\dagger} = U \sqrt{Y^\dagger Y}U^\dagger
$$
 and squaring both sides of the preceding equation yields
$$
YY^\dagger = U Y^\dagger YU^\dagger.
$$
Thus $YY^\dagger$ and $Y^\dagger Y$ are unitarily equivalent and thus they have the same eigenvalues counting multiplicities.  In fact,  suppose that $\lambda$ is an eigenvalue for $YY^\dagger$ with corresponding eigenvector $\v$.  Then
\begin{eqnarray*}
 0 = (YY^\dagger - \lambda I)\v &=& ( U Y^\dagger YU^\dagger-  \lambda I)\v \\
 							& = & U(Y^\dagger Y- \lambda I)U^\dagger\v.
							\end{eqnarray*}
We see $0 = U(Y^\dagger Y - \lambda I)U^\dagger\v$ and multiplying both sides on the left by $U^\dagger$ yields
$$
 0 = (Y^\dagger Y - \lambda I)U^\dagger\v.
$$
Thus $U^\dagger\v$ is an eigenvector for  $Y^\dagger Y$ with eigenvalue $\lambda$.    Bottom line: { $Y^\dagger Y$ and $YY^\dagger$ always have the same eigenvalues counting multiplicities.}  We apply this fact below to the pairs  $E^\dagger E$ and $EE^\dagger$ and $W^\dagger W$ and $WW^\dagger$.
 
Eq.\  (\ref{NDN}) shows us that the set of eigenvalues of $E^\dagger E$ is
 $$
 \{b_0/x, b_1/x, \ldots, b_{d-1}/x\}
 $$
 while (\ref{WDW}) shows us that the set of eigenvalues of $W^\dagger W$ is
\begin{equation}\label{WEV}
 \left\{\frac{1-b_0}{1-x}, \frac{1-b_1}{1-x}, \ldots, \frac{1-b_{d-1}}{1-x}\right\}.
\end{equation}
Since for $j=0, 1, 2, \ldots, d-1$,  $b_j/x$ is an eigenvalue of $E^\dagger E$, it is also an eigenvalue of $EE^\dagger$.   Let $\v_j$ be an eigenvector for $EE^\dagger$ with corresponding eigenvalue $\frac{b_j}{x}$.   Applying both sides of (\ref{RowKey}) to $\v_j$ and doing a bit of rearranging, we obtain
 $$
 WW^\dagger\v_j = \left(\gamma - \frac{b_j}{x}\right)\v_j.
 $$
 Thus $\left(\gamma - \frac{b_j}{x}\right)$ is an eigenvalue of $WW^\dagger$.   In fact, it is easy to see the set of eigenvalues of $WW^\dagger$ is precisely $\{\gamma - \frac{b_0}{x}, \gamma - \frac{b_1}{x}, \ldots, \gamma - \frac{b_{d-1}}{x}\}$.

 From Eq.\ (\ref{EKE}), we have
 $$
d^2 - 2+\frac{b_j}{x} + \frac{1-b_j}{1-x} =  \frac{d}{\lambda_j},
 $$
 so that 
\begin{equation}\label{bje}
 b_j = -\frac{x}{1-2x} + \frac{x(1-x)}{1-2x}\left(\frac{d}{\lambda_j} +2 -  d^2\right).
\end{equation}
 Because the sequence $\lambda_j$ decreases with $j$, the preceding equation shows that $b_j$ increases with $j$:
\begin{equation}\label{binc}
  b_0 \le b_1\le \cdots \le b_{d-1}.
\end{equation}

Thus,  the work of the preceding paragraph shows that 
 $$
 \gamma - \frac{b_0}{x}
 $$
 must be the largest eigenvalue of $WW^\dagger$; equivalently, the largest eigenvalue of $W^\dagger W$ (since $WW^\dagger$ and $W^\dagger W$ share the same eigenvalues).  Hence, recalling our earlier listing (\ref{WEV}) of the eigenvalues  and (\ref{binc}), we must have 
 $$
 \frac{1-b_0}{1-x} = \gamma - \frac{b_0}{x}.
 $$
Rearranging and using $\gamma = \sum_{j=0}^{d-1}\frac{1}{\lambda_j} - (d^2-2)$,  we see that the preceding equation yields
 $$
 \sum_{j=0}^{d-1}\frac{1}{\lambda_j} - (d^2-2) = \frac{1-b_0}{1-x} +\frac{b_0}{x}.
 $$
 Now substitute the left-hand side of the preceding equation for the right-hand side in Eq.\  (\ref{EKE}) in the $j =0$ case:
 $$
\lambda_0\left[ d^2 - 2 + \left(\sum_{j=0}^{d-1} \frac{1}{\lambda_j} -(d^2-2)\right)\right]=d,
 $$
 so that
 $$
\sum_{j=0}^{d-1} \frac{\lambda_0}{\lambda_j}=d.
$$
 We know that  $\lambda_0 \ge \lambda_j$ for $j =0, 1, 2, \ldots, d-1$, so that $\lambda_0/\lambda_j\ge 1$ for each $j$.  From the preceding equation, we conclude that $\lambda_0/\lambda_j = 1$ for all $j$, i.e., all Schmidt coefficients of $\ket{\Psi}$ are equal, which means $\ket{\Psi}$ must be maximally entangled, as was to be proved.

\section{Alice can encode $d^2-1$ messages with high probability for certain non-maximally entangled states}  

We continue to assume that Alice and Bob share a two-qudit system in state $\ket{\Psi}$ of (\ref{GQDS}) with largest Schmidt coefficient $\sqrt{\lambda_0}$.   In this section, we show that for certain non-maximally entangled states (those for which $\lambda_0$ is small but still exceeds $1/d$), Alice can encode $d^2-2$ perfectly distinguishable messages via original-protocol unitary encoding and, with high probability, can use a non trace-preserving quantum operation to encode a $(d^2-1)$-th message perfectly distinguishable from those encoded according to the original protocol.  

 We assume {\em throughout this section} that $\ket{\Psi}$ of (\ref{GQDS})  is not maximally entangled ($\lambda_0 > 1/d$) yet it is entangled enough to permit Alice to send to Bob $d^2 - 2$ perfectly distinguishable messages via unitary encoding operators $U^{(n)}$ on $H_A$, $n = 0, 1, \ldots, d^2 -3$.   The result  (\ref{WCSGB}) of Wu {\it et al.}\  shows that $\lambda_0 \le \frac{d}{d^2-2}$.  We will assume that  $\lambda_0$ is strictly less than $ d/(d^2-2)$ and hence $\lambda_j < d/(d^2-2)$ for $j=0, 1, 2,\ldots, d-1$.  It follows that  all Schmidt coefficients of $\ket{\Psi}$ must be nonzero (consistent with our standing assumption); in addition, we have for all $j$,
\begin{equation}\label{RJ}
R_j \equiv d - (d^2 - 2)\lambda_j
\end{equation}
is positive.     Note well that $R_j$ increases with $j$: $R_0\le R_1\le \cdots\  \le R_{d-1}$.

We continue to assume that the natural basis $\B$ for the  Alice-Bob system is ordered as in (\ref{Bwo}).

  We construct a $d^2\times d^2$ matrix $M$ as follows.   Just as in the preceding section,  the coordinate vectors of $(U^{(n)}\otimes I_B)\ket{\Psi}$, $n = 0, 1, \ldots, d^2-3$,  relative to the basis $\B$,  give the first $d^2-2$ columns of $M$.  By adding two appropriately chosen columns, we  can extend these $d^2-2$ orthonormal columns to form a $d^2\times d^2$ {\em unitary} matrix $M$.   Let $\v$ be the penultimate column and $\w$ be the final column of this unitary matrix $M$.   Label the entries of $\v$ and $\w$ according to the ordering of $\B$ (just as we labeled the entries of the column vectors for $\ket{\phi_0}/\sqrt{x}$ and $\ket{\phi_1}{\sqrt{1-x}}$ in the preceding section) so that, e.g., the entries of $\v$ are $v_{00}, v_{10}, v_{20}, \ldots , v_{d-1,0}, v_{01}, v_{11}, \ldots, v_{d-1,1}, \ldots, v_{d-1,d-1}$.

  Form a $d\times d$ matrix $T$ such that
  $$
  t_{ij} = \frac{\sqrt{\lambda_{d-1}}}{\sqrt{\lambda_j}\sqrt{R_{d-1}}} v_{ij},
  $$
  where $R_j$ is defined by (\ref{RJ}).
  For example, if $d = 3$, we have
  $$
 T = \begin{bmatrix} \frac{\sqrt{\lambda_2}}{\sqrt{R_{2} \lambda_0}} v_{00}& \frac{\sqrt{\lambda_2}}{\sqrt{R_{2}\lambda_1}}v_{01} & \frac{1}{\sqrt{R_{2}}}v_{02}\\  \frac{\sqrt{\lambda_2}}{\sqrt{R_{2}\lambda_0}} v_{10}& \frac{\sqrt{\lambda_2}}{\sqrt{R_{2}\lambda_1}}v_{11} & \frac{1}{\sqrt{R_{2}}}v_{12}\\ \frac{\sqrt{\lambda_2}}{\sqrt{R_{2} \lambda_0}} v_{20}& \frac{\sqrt{\lambda_2}}{\sqrt{R_{2}\lambda_1}}v_{21} & \frac{1}{\sqrt{R_{2}}}v_{22}\end{bmatrix}.
  $$
  Similarly, define the $d\times d$ matrix $Y$ by
  $$
 y_{ij} =  \frac{\sqrt{\lambda_{d-1}}}{\sqrt{\lambda_j}\sqrt{R_{d-1}}} w_{ij}.
  $$

Claim:
  $$
 T^\dagger T + Y^\dagger Y \le I, \quad {\rm i.e.} \quad I - T^\dagger T - Y^\dagger Y \quad {\rm is\  a\ positive\ matrix}.
  $$
(Here $I$ is the $d\times d$ identity matrix.) We now justify this claim.   Using the structure of the matrix $M$,  we sum the squared magnitudes of rows $0,j$ through $d-1,j$ to get 
$$
(d^2-2)\lambda_j + \sum_{i=0}^{d-1}|v_{ij}|^2 + \sum_{i=0}^{d-1} |w_{ij}|^2 = d,
$$
where we have used the facts that the rows of $M$ have length one and that the entries $0,j$ through $d-1,j$ in any one of the first $d^2-2$ columns of $M$ constitute the column of a unitary matrix.
Thus, 
\begin{equation}
\sum_{i=0}^{d-1}|v_{ij}|^2 + \sum_{i=0}^{d-1} |w_{ij}|^2 = d-  (d^2-2)\lambda_j = R_j.
\end{equation}
Note that for $j = 0, 1, \ldots, d-1$,
$$
(T^\dagger T + Y^\dagger Y)_{jj} = \frac{\lambda_{d-1}}{\lambda_j}\frac{1}{R_{d-1}}\left(\sum_{i=0}^{d-1}|v_{ij}|^2 + \sum_{i=0}^{d-1} |w_{ij}|^2\right) =  \frac{\lambda_{d-1}}{\lambda_j}\frac{R_j}{R_{d-1}}.
$$
Since $\lambda_j$ decreases with $j$ while $R_j$ increases with $j$ and both are positive, we see that 
$$
1 \ge (T^\dagger T + Y^\dagger Y)_{jj} > 0
$$ 
for each $j$; moreover,
$$
1 = (T^\dagger T + Y^\dagger Y)_{d-1, d-1}.
$$

 Thus the diagonal entries of $I - T^\dagger T - Y^\dagger Y$ are nonnegative with the final entry being $0$.  We assert that the off-diagonal entries of $I - T^\dagger T - Y^\dagger Y$ are all zeros, equivalently, that the off-diagonal entries of $T^\dagger T +  Y^\dagger Y$  are all zeros.  Let $r$ and $s$ be distinct elements of $\{0, 1, \ldots, d-1\}$.   Because the matrix $M$ is unitary, the inner product of each pair of distinct rows of $M$ is $0$.  Thus the inner product of the $ir$ and $is$ rows of the matrix $M$ vanishes,  as therefore does the sum over $i$  of these inner products (where we are assuming 
of course that $r \ne s$ ). Accordingly, recalling our explanation earlier in this  section of how the matrix $M$ is constructed, we obtain
\begin{eqnarray*}
 0 & = & \sum_{i=0}^{d-1} \left(\sqrt{\lambda_r\lambda_s} \sum_{n=0}^{d^2-3} (u^{(n)}_{ir})^*u^{(n)}_{is}\right) + \sum_{i=0}^{d-1} v_{ir}^*v_{is} + \sum_{i=0}^{d-1}w_{ir}^*w_{is}\\
  & = & \sqrt{\lambda_r\lambda_s}  \sum_{n=0}^{d^2-3} \sum_{i=0}^{d-1} (u^{(n)}_{ir})^*u^{(n)}_{is} + \sum_{i=0}^{d-1} v_{ir}^*v_{is} + \sum_{i=0}^{d-1}w_{ir}^*w_{is}\\
  & = & \sum_{i=0}^{d-1} v_{ir}^*v_{is} + \sum_{i=0}^{d-1}w_{ir}^*w_{is},
  \end{eqnarray*}
  where the final equality holds because for each $n\in \{0, 1, 2, \ldots, d^2-3\}$, the matrix $U^{(n)}$ is unitary (in particular, its columns are orthogonal).  We have $ 0 =  \sum_{i=0}^{d-1} v_{ir}^*v_{is} + \sum_{i=0}^{d-1}w_{ir}^*w_{is}$ so that
 \begin{eqnarray*}
 0 &= & \frac{\lambda_{d-1}}{R_{d-1}\sqrt{\lambda_r\lambda_s}}\left(  \sum_{i=0}^{d-1} v_{ir}^*v_{is} + \sum_{i=0}^{d-1}w_{ir}^*w_{is}\right)\\ 
   & = &  \left(T^\dagger T+ Y^\dagger Y\right)_{rs},
  \end{eqnarray*}
  as desired.

  We have shown $I - T^\dagger T- Y^\dagger Y$ is positive:  it's a diagonal matrix with nonnegative entries along the diagonal.  
  
We view $T$ and $Y$ as operators on $H_A$.    It is easy to see that  $(T\otimes I_B)\ket{\Psi}$ and $(Y\otimes I_B)\ket{\Psi}$ are each orthogonal to the messages $(U^{(n)}\otimes I_B)\ket{\Psi}$, $n = 0, 1, \ldots, d^2-3$.  The coordinate vector of $(T\otimes I_B)\ket{\Psi}$ relative to $\B$ is just $\frac{\sqrt{\lambda_{d-1}}}{\sqrt{R_{d-1}}}\v$ and that for $(Y\otimes I_B)\ket{\Psi}$ is $\frac{\sqrt{\lambda_{d-1}}}{\sqrt{R_{d-1}}}\w$.  Since $\v$ and $\w$  are each orthogonal to the first $d^2-2$ columns of $M$, so are these scalar multiples of $\v$ and $\w$.  Note that neither of the vectors
$\frac{\sqrt{\lambda_{d-1}}}{\sqrt{R_{d-1}}}\v$ or $\frac{\sqrt{\lambda_{d-1}}}{\sqrt{R_{d-1}}}\w$ is the zero vector (so that neither $T$ nor $Y$ is the zero matrix; in fact, it is easy to see $T$ and $Y$ are linearly independent).

   Define $C$ to be the square root of  $I - T^\dagger T - Y^\dagger Y$.  Note $C$ is a diagonal matrix whose diagonal entries are square roots of the diagonal entries of $I - T^\dagger T - Y^\dagger Y$.   We have
\begin{equation}\label{ABCK}
  T^\dagger T + Y^\dagger Y + C^\dagger C = I.
\end{equation}
We have already noted that $(C^\dagger C)_{d-1,d-1} = 0$. Now,  for $j = 0, 1, \ldots, d-2$, we have
\begin{eqnarray}
  (C^\dagger C)_{jj}  &= &1- \frac{\lambda_{d-1}}{\lambda_j}\frac{R_j}{R_{d-1}} \nonumber \\
                                        & = & \frac{d(\lambda_j- \lambda_{d-1})}{\lambda_j[d-(d^2-2)\lambda_{d-1}]} \label{IURS}\\
                                        &\le &\frac{d^2(\lambda_j- \lambda_{d-1})}{2\lambda_j}, \label{EE}
                                 \end{eqnarray}
where to obtain the second equality we have used Eq.\ (\ref{RJ}) and to obtain the final inequality we have used $\lambda_{d-1} \le 1/d$.  Since $\lambda_0 \ge \lambda_j \ge \lambda_{d-1} \ge 1- (d-1)\lambda_0$, both $\lambda_j$ and $\lambda_{d-1}$ approach $1/d$ as $\lambda_0$ approaches $1/d$.  Thus,  the overestimate (\ref{EE}) for $(C^\dagger C)_{jj}$ shows that all diagonal entries of $C^\dagger C$ approach $0$  as $\lambda_0$ approaches $1/d$.      For $j = 0, 1, \ldots, d-1$, let $\gamma_j = (C^\dagger C)_{jj}$, so that $C$ is a $d\times d$ diagonal matrix with diagonal entries $\sqrt{\gamma_0}, \ldots, \sqrt{\gamma_{d-1}}$, and, as we just discussed, for each $j$, $\gamma_j\rightarrow 0$ as $\lambda_0 \rightarrow 1/d$.

  Because $T$, $Y$, and $C$ satisfy the ``Kraus-operator condition'' (\ref{ABCK}), a process, described on p.\ 365 of \cite{NC}, e.g., establishes that  Alice can pair her qudit with an ancillary qutrit $a$  and perform a unitary operator $\tilde{U}$ on the pair
 to cast the $ABa$ triple, in initial state $\ket{\Psi}\ket{0}_a$, into the state (\ref{ROU}) with $N=3$, $K^{(0)} = T$, $K^{(1)} = Y$, and $K^{(2)} = C$:
\begin{equation}\label{WYC}
   (T \otimes I_B )\ket{\Psi}\otimes \ket{0}_a +   (Y \otimes I_B)\ket{\Psi}\otimes \ket{1}_a +   (C \otimes I_B)\ket{\Psi}\otimes \ket{2}_a.
\end{equation}
The corresponding reduced-density operator description of the resulting state of the $AB$ system is $\E(\ketbra{\Psi}{\Psi})$, where
\begin{equation}\label{ETY}
 \E(\rho) = T\rho T^\dagger +Y\rho Y^\dagger + C\rho C^\dagger.
\end{equation}
Because we are assuming that $\ket{\Psi}$ is not maximally entangled,  $C$ cannot be the zero matrix.  If it were, this would contradict the work of Section~\ref{CBJ}, because Alice could then use $\E$  of (\ref{ETY}) to encode a $(d^2-1)$-th message for Bob perfectly distinguishable from the unitary messages represented by $(U^{(n)}\otimes I_B)\ket{\Psi}$, $n=0, 1, \ldots, d^2-3$ (because $(T\otimes I_B)\ket{\Psi}$ and $(Y\otimes I_B)\ket{\Psi}$ are orthogonal to the ``unitary messages'').  

We now show that under appropriate conditions, Alice's measurement of the ancilla $a$ after applying $\tilde{U}$ to the $Aa$ pair, can, with high probability, create a $(d^2-1)$-th message for Bob perfectly distinguishable from the initial $d^2-2$ messages.    Alice and Bob can agree (say, before they part company) that Bob will decode messages from Alice via the observable 
\begin{equation}\label{BOBL}
\sum_{j=0}^{d^2-2} j P_{S_j}
\end{equation}
where, for $j = 0, 1, \ldots, d^2-3$, $S_j$ is the one dimensional subspace of $H$ spanned by $(U^{(j)}\otimes I_B)\ket{\Psi}$; and, for $j = d^2-2$, $S_j$ is the two dimensional subspace on $H$ spanned by $(T\otimes I_B)\ket{\Psi}$ and $(Y\otimes I_B)\ket{\Psi}$ (and $P$ stands for projection).  Since the subspaces $S_j$ are pairwise orthogonal, Bob will receive perfectly distinguishable messages as long as Alice either encodes via some selected one of the original-protocol unitaries, or else encodes via a quantum
operation (in this case non trace-preserving)  that yields a state of the $AB$ system described by a density operator whose support is $S_{d^2-2}$.   Alice has no trouble producing a unitary message by applying a unitary operation to her qudit alone.  She can thereby  produce $d^2-2$ perfectly distinguishable messages.  To (attempt to) produce the final $(d^2-1)$-th message, Alice applies the unitary operation $\tilde{U}$ to the qudit-qutrit pair $Aa$, as described in the preceding paragraph, casting the $ABa$ system  into the state (\ref{WYC}).  Then she performs the projective measurement on $ABa$ corresponding to $P_{\ket{2}_a} \equiv (I_A\otimes I_B \otimes \ket{2}_a\bra{2}_a)$; she will measure $1$ (i.e., the state $\ket{2}_a$) with probability
\begin{eqnarray}
  p_1 &=& \left\|  (C \otimes I_B)\ket{\Psi}\right\|^2 \nonumber\\
  & = & \left\| \sum_{j=0}^{d-1} \sqrt{\lambda_j}\sqrt{\gamma_j}\ket{jj}\right\|^2\nonumber\\
  & = & \sum_{j=0}^{d-1}\lambda_j\gamma_j,\label{EVPR}
  \end{eqnarray}
where the second equality follows from the fact that $C$ is a diagonal matrix (with diagonal entries $\sqrt{\gamma_j}$\ ) and where we use $\| \cdot \|$ to denote vector length.   Upon substituting our overestimate (\ref{EE}) for $\gamma_j$ into (\ref{EVPR}) and using $\sum_{j=0}^{d-1} \lambda_j = 1$, we obtain  $p_1 \le (d^2/2) - (d^3/2)\lambda_{d-1}$, which,  by applying  $\lambda_{d-1} \ge 1- (d-1)\lambda_0$,  yields
\begin{equation}\label{WRWT}
p_1 \le \frac{d^3(d-1)}{2} \left(\lambda_0 - \frac{1}{d}\right).
\end{equation}
  Thus for $\lambda_0$ sufficiently close to $1/d$ the probability $p_1$ that Alice will measure $1$ via the projective measurement $P_{\ket{2}_a}$ approaches $0$.   Thus, with probability $1-p_1$ (approaching $1$ as $\lambda_0 \rightarrow 1/d$),  the measurement  $P_{\ket{2}_a}$ will yield $0$, casting the $ABa$ system into the state
\begin{equation}\label{DS}
\frac{1}{\sqrt{1-p_1}}\left[  \vstrut (T \otimes I_B )\ket{\Psi}\otimes \ket{0}_a +   (Y \otimes I_B)\ket{\Psi}\otimes \ket{1}_a\right].
\end{equation}
  At this point Alice can send her qudit to Bob (through a noiseless quantum channel) and using his observable modeled by (\ref{BOBL}), Bob will receive message $d^2-1$ with certainty.   
   
  Observe that the reduced density operator corresponding to (\ref{DS}) for the state of the $AB$ system is
  $$
  \frac{1}{1-p_1} \left[  \vstrut (T \otimes I_B )\ketbra{\Psi}{\Psi} (T \otimes I_B)^\dagger +   (Y \otimes I_B)\ketbra{\Psi}{\Psi} (Y \otimes I_B)^\dagger \right],
  $$
which is associated with  the non trace-preserving (because $C\ne 0$) quantum operator 
\begin{equation}\label{NTPQO}
\rho\mapsto (T \otimes I_B )\rho (T \otimes I_B)^\dagger +   (Y \otimes I_B)\rho (Y \otimes I_B)^\dagger.
\end{equation}
  Thus with probability $1-p_1$, Alice can use the non trace-preserving quantum operator defined by (\ref{NTPQO}) to encode a $(d^2-1)$-th message for Bob, and he will never mistake this message for any of the original-protocol unitary messages she may encode.   A nice feature of this dense-coding scheme, which allows Alice to send $d^2-1$ perfectly distinguishable messages to Bob with high probability (and the first $d^2-2$ of those messages with certainty), is that if Alice does  measure the ancilla to be in the undesirable state $\ket{2}_a$, then she can choose not to send her qudit to Bob since he would not be assured of receiving her intended message. If Alice does wish to send the $(d^2-1)$-th message to Bob, we have shown that the probability of failure $p_1$ of the protocol has the upper bound (\ref{WRWT}), which gives an indication of how $p_1$ decreases to $0$ as $\lambda_0$ approaches $1/d$.    We note that the overestimate (\ref{WRWT})  of $p_1$ does not typically provide a sharp bound on the probability of failure.  Suppose, e.g.,  all Schmidt coefficients of  $\ket{\Psi}$, except the largest, are equal, implying $\lambda_j= \lambda_{d-1} = (1-\lambda_0)/(d-1)$ for $j = 1, 2, \ldots, d-2$.   Then, recalling $(C^\dagger C)_{jj}  = \gamma_j$, we see that  Eq.\  (\ref{IURS}) shows that $\gamma_j = 0$ for $j=1, 2, \ldots, d-1$; thus  (\ref{IURS})  combined with (\ref{EVPR}) yields  
 \begin{equation}\label{NOEE}
  p_1 = \frac{d^2(\lambda_0 - 1/d)}{d^2(\lambda_0 - 1/d) + 2(1-\lambda_0)} \le \frac{d^3}{2(d-1)}\left(\lambda_0 - \frac{1}{d}\right),
\end{equation}
where the inequality holds because  $d^2(\lambda_0 - 1/d) + 2(1-\lambda_0)$ increases with $\lambda_0$, taking the value $2(d-1)/d$ when $\lambda_0 = 1/d$.   Because $\lambda_0$ cannot exceed $d/(d^2 - 2)$ (as explained at the opening of this Section), the  bound on the probability of failure given by (\ref{NOEE}) cannot exceed $d^2/[(d - 1)(d^2 - 2)]$, which decreases asymptotically with increasing $d$Ê as $1/d$.   If we assume that $\lambda_0 = d/(d^2-1)$ (the value of the bound (\ref{WCSGB}) for $L = d^2-1$),  so that $\lambda_0- 1/d$ is roughly 50 percent of its maximum allowed value, then the bound (\ref{NOEE}) yields small error probabilities even for small $d$; for example 28\% for d =3, dropping to 8.5\% for $d = 7$.

  We conclude with a concrete example presenting a situation in which Alice has over a 97\% chance of communicating three perfectly distinguishable messages to Bob using a system of two less than maximally entangled qubits.    Suppose Alice and Bob share a two-qubit system in state
\begin{equation}\label{TQB}
 \ket{\Psi} =  \frac{9}{4\sqrt{10}} \ket{00} + \frac{\sqrt{79}}{4\sqrt{10}}\ket{11},
\end{equation}
  for which $\lambda_0 = 81/160$ and $\lambda_1 = 79/160$. 
  For the preceding state, Alice may choose unitary encoding operations on her qubit $A$ corresponding to the identity $I_A = \ketbra{0}{0} + \ketbra{1}{1}$ and the shift operator $X= \ketbra{1}{0} + \ketbra{0}{1}$.   Here's a matrix $M$ for this situation, whose first two columns, respectively,  are the coordinates of $(I_A\otimes I_B)\ket{\Psi}$ and $(X\otimes I_B)\ket{\Psi}$ relative to $\B = (\ket{00},\ket{10},\ket{01},\ket{11})$ and whose remaining columns extend the first two to an orthonormal basis of $\C^4$:

  $$
  M =\left[ \begin {array}{cccc} {\frac {9}{40}}\,\sqrt {10}&0&\frac{1}{40}\,\sqrt {395}&\frac{1}{40}\,\sqrt {395}\\\noalign{\medskip}0&{\frac {9}{40}}\,\sqrt {10}&\frac{1}{40}\,\sqrt {395}&-\frac{1}{40}\,\sqrt {395}\\\noalign{\medskip}0&\frac{1}{40}\,\sqrt {79}\sqrt {10}&-{\frac {9}{40}}\,\sqrt {5}&{\frac {9}{40}}\,\sqrt {5}\\\noalign{\medskip}\frac{1}{40}\,\sqrt {79}\sqrt {10}&0&-{\frac {9}{40}}\,\sqrt {5}&-{\frac {9}{40}}\,\sqrt {5}\end {array} \right]  .
  $$

     The matrices $T$, $Y$, and $C$ are
  $$
T=\left[ \begin {array}{cc} {\frac {79}{162}}&-1/2\\\noalign{\medskip}{\frac {79}{162}}&-1/2\end {array} \right] ,  Y = \left[ \begin {array}{cc} {\frac {79}{162}}&1/2\\\noalign{\medskip}-{\frac {79}{162}}&-1/2\end {array} \right],  C = \left[ \begin {array}{cc} {\sqrt{\frac {320}{6561}}}&0\\\noalign{\medskip}0&0\end {array} \right].
 $$

 Note $\gamma_0$ the upper left entry of $C^\dagger C$ is $320/6561$ while $\gamma_1$ the lower right entry is $0$ (no surprise).   
The probability that Alice will measure the ancillary qutrit $a$ to be in state $\ket{2}_a$ after applying the unitary $\tilde{U}$ to the $Aa$  system (as described above) is
$\lambda_0 \gamma_0 = 81/160 \cdot 320/6561 = 2/81$.  Thus the probability that this measurement will cast the system into the desirable state (\ref{DS}) is $79/81 \approx 97.5 \%$.  Thus there is over an 97\% chance that Alice and Bob can use the less-than-maximally-entangled  state (\ref{TQB}) as a resource for the communication of 3 perfectly distinguishable messages.  Specifically, Alice can communicate ``message 0'',  generated by applying $I_A$ to $A$, with certainty; ``message 1'', generated by applying $X$ to $A$, with certainty; and  ``message 2'',  generated by applying $\tilde{U}$ to $Aa$ and then measuring $a$, with probability $79/81$.  If in attempting to encode message 2, Alice applies $\tilde{U}$ to $Aa$ and then observes $a$ to be in the undesirable state $\ket{2}_a$, then the state of the system she shares with Bob is $(C\otimes I_B)\ket{\Psi}/\|C\otimes I_B)\ket{\Psi}\| = \ket{00}$, and she should not  send her qubit to Bob since he would misinterpret her intended message (message 2) as message 0.  

If Alice and Bob are willing to tolerate a small chance of Bob's misinterpreting a message, they will be better served if Alice encodes message 2 by simply applying $\tilde{U}$ to $Aa$  and then sending her qubit  to Bob {\em without  measuring the ancilla $a$}.     In the example above, the projectors that Bob would use for decoding the message that Alice sends are
$$
P_{S_0} = \ketbra{\Psi}{\Psi},
$$
$$
P_{S_1} =  (X\otimes I_B)\ket{\Psi}\bra{\Psi}(X\otimes I_B)^\dagger,
$$
and
$$
P_{S_2} = \frac{1}{p_T} (T\otimes I_B)\ket{\Psi}\bra{\Psi}(T\otimes I_B)^\dagger + \frac{1}{p_Y} (Y\otimes I_B)\ket{\Psi}\bra{\Psi}(Y\otimes I_B)^\dagger,
$$
where $p_T = \|(T\otimes I_B)\ket{\Psi}\|^2 = 79/162$ and $p_Y = \|(Y\otimes I_B)\ket{\Psi}\|^2 = 79/162$.
Suppose that Alice performs only the unitary $\tilde{U}$ on the the $Aa$ qubit-qutrit system {\em and does not measure the ancilla}.  If Alice then sends her qubit to Bob (representing transmission of ``message 2''), then  Bob would measure $2$ with probability:
\begin{equation*}
\begin{split}
\tr\left(P_{S_2} (T\otimes I_B)\ket{\Psi}\bra{\Psi}(T\otimes I_B)^\dagger  + P_{S_2}  (Y\otimes I_B)\ket{\Psi}\bra{\Psi}(Y\otimes I_B)^\dagger  + P_{S_2}  (C\otimes I_B)\ket{\Psi}\bra{\Psi}(C\otimes I_B)^\dagger \right) \\ =\tr\left((T\otimes I_B)\ket{\Psi}\bra{\Psi}(T\otimes I_B)^\dagger  + (Y\otimes I_B)\ket{\Psi}\bra{\Psi}(Y\otimes I_B)^\dagger  + P_{S_2}  (C\otimes I_B)\ket{\Psi}\bra{\Psi}(C\otimes I_B)^\dagger \right) \\ = 79/80 =98.75\%.
\end{split}
\end{equation*}
Bob will always decode messages $0$ and $1$ correctly.      

\acknowledgements

The authors are indebted to Scott Cohen, Vlad Gheorghiu, Robert Griffiths, and Thomas Williams for useful discussions.

\appendix
\section{Proof of Lemma II.1}
 \begin{lem21}  Suppose that all the Schmidt coefficients of $\ket{\Psi}$ in the representation (\ref{GQDS}) are nonzero and that  $K^{(0)}, K^{(1)}, \ldots, K^{(m-1)}$ are linearly independent in $\L(H_A)$; then $(K^{(0)}\otimes I)\ket{\Psi}, (K^{(1)}\otimes I)\ket{\Psi}, \ldots, (K^{(m-1)}\otimes I)\ket{\Psi}$ are linearly independent vectors in $H$.
  \end{lem21}
  \begin{proof}  Suppose that $\alpha_0, \alpha_1, \ldots, \alpha_{m-1}$ are scalars such that
    $$
   \sum_{p=0}^{m-1} \alpha_p (K^{(p)}\otimes I)\ket{\Psi} = 0.
   $$
   Then, letting $k_{ij}^{(p)} = \bra{i}K^{(p)}\ket{j}$, we obtain
   $$
 \sum_{i,j=0}^{d-1}\left( \sqrt{\lambda_j} \sum_{p=0}^{m-1} \alpha_p k_{ij}^{(p)}\right) \ket{ij} = 0, 
   $$
   which implies, upon taking the inner product of both sides of the preceding equation with $\ket{ij}$,  that for each $i$ and $j$
\begin{equation}\label{FE}
  \sum_{p=0}^{m-1} \alpha_p k_{ij}^{(p)} =0
\end{equation}
(since $\sqrt{\lambda_j}$ is nonzero for each $j$).    Note that (\ref{FE}) says
$$
\sum_{p=0}^{m-1} \alpha_p K^{(p)} =0,
$$
which implies $\alpha_p = 0$ for all $p$ since  $K^{(0)}, K^{(1)}, \ldots, K^{(m-1)}$ are   linearly independent.   Thus, $(K^{(0)}\otimes I)\ket{\Psi}, (K^{(1)}\otimes I)\ket{\Psi}, (K^{(m-1)}\otimes I)\ket{\Psi}$ are  linearly independent.

\end{proof}
  \section{Kraus operator representations}

     Suppose that Alice pairs her qudit $A$ with an ancillary $N$-level particle $a$.   The Hilbert space for the $Aa$ pair is $H_A\otimes H_a$ with orthonormal basis $\B_{Aa} \equiv \{\ket{i}_A\ket{r}_a: i = 0, 1, 2, \ldots, d-1; r=0, 1, 2, \ldots, N-1\}$.   Assume, that Alice's particle $A$ is entangled with Bob's particle $B$ and their two-qudit system is in state $\ket{\Psi}$ given by (\ref{GQDS}).   Assume that $a$ is in state $\ket{0}_a$.   Suppose the physical equivalent of a unitary operator $U$ on $H_A \otimes H_a$ is applied to the pair $Aa$.      
    
     We can express $U$ in terms of its action on the the basis elements in $\B_{Aa}$ as follows
\begin{equation}\label{Urep}
\begin{split}
    U = \sum_{i,j=0}^{d-1}\sum_{r,s=0}^{N-1} u_{ir, js} \ket{i}_A\ket{r}_a \bra{j}_A \bra{s}_a\\
      =  \sum_{i,j=0}^{d-1}\sum_{r,s=0}^{N-1}  u_{ir, js}\left(\vstrut  \ket{i}_A\bra{j}_A \otimes \ket{r}_a \bra{s}_a\right),
      \end{split}
      \end{equation}
      where the scalars $u_{ir,js}$ must satisfy the following condition owing to the unitarity of $U$: for $0 \le j,j' \le d-1$ and $0\le s,s'\le N-1$,
  \begin{equation}\label{UP}
  \sum_{i=0}^{d-1}\sum_{r=0}^{N-1} u_{ir, js}^*u_{ir,j's'} = \delta_{j,j'}\delta_{s,s'}.
 \end{equation}
    
Letting $I_B$ be the identity on Bob's Hilbert space $H_B$, we see that the effect of Alice's unitary operator $U$ on the state $\ket{\Psi}\ket{0}_a$ may be described as follows.
\begin{eqnarray*}
\left(\sum_{i,j=0}^{d-1}\sum_{r,s=0}^{N-1}  u_{ir, js}\left(\vstrut  \ket{i}_A\bra{j}_A\otimes I_B\right) \otimes \left(\ket{r}_a \bra{s}_a\right) \right)\ket{\Psi}\otimes\ket{0}_a
 & = &  \sum_{i,j=0}^{d-1}\sum_{r,s=0}^{N-1}  u_{ir, js} \left(\vstrut \ket{i}_A\bra{j}_A \otimes I_B\right)\ket{\Psi}\otimes (\ket{r}_a \bra{s}_a)\ket{0}_a \nonumber \\
 &=& \sum_{r=0}^{N-1}  \sum_{i,j=0}^{d-1} u_{ir, j0} \left(\vstrut \ket{i}_A\bra{j}_A\otimes I_B \right)\ket{\Psi}\otimes \ket{r}_a \nonumber \\
 &=& \sum_{r=0}^{N-1} \left(\left(\vstrut \sum_{i,j=0}^{d-1} u_{ir, j0}  \ket{i}_A\bra{j}_A\right)\otimes I_B \right)\ket{\Psi}\otimes \ket{r}_a\nonumber\\
 &=&  \sum_{r=0}^{N-1} \left(  K^{(r)} \otimes I_B \right)\ket{\Psi}\otimes \ket{r}_a,
 \end{eqnarray*}
 where $K^{(r)} = \sum_{i,j=0}^{d-1} u_{ir, j0}  \ket{i}_A\bra{j}_A$ are Kraus operators that satisfy $\sum_{r=0}^{N-1}  (K^{(r)})^\dagger K^{(r)} = I_A$:
 \begin{eqnarray*}
\sum_{r=0}^{N-1}  (K^{(r)})^\dagger K^{(r)} &=& \sum_{r=0}^{N-1} \left(\sum_{i,j=0}^{d-1} u_{ir, j0}^* \ket{j}_A\bra{i}_A\right)\left(\sum_{i',j'=0}^{N-1} u_{i'r,j'0} \ket{i'}_A\bra{j'}_A\right)\\
  & = & \sum_{r=0}^{N-1} \sum_{i,j,j'=0}^{d-1} u_{ir,j0}^*u_{ir,j'0} \ket{j}_A\bra{j'}_A \\
  & = &  \sum_{j,j'=0}^{d-1}  \sum_{i=0}^{d-1}\sum_{r=0}^{N-1} u_{ir, j0}^*u_{ir,j'0}\ket{j}_A\bra{j'}_A \\
  &=& \sum_{j,j'=0}^{d-1} \delta_{j,j'} \ket{j}_A\bra{j'}_A\\
  & = &  \sum_{j=0}^{d-1} \ket{j}_A\bra{j}_A\\
  & = & I_{A},
  \end{eqnarray*}
  where we have used (\ref{UP}) to obtain the fourth equality above.    Thus the process of pairing $A$ with the $N$-level particle $a$ and applying $U$ to the pair always casts the three-particle system $ABa$ into a state  of the form (\ref{ROU}), where the Kraus operators $K^{(0)}, \ldots, K^{(N-1)}$ satisfy  (\ref{TPC}).    The quantum operator $\E$ associated with this process will have the form (\ref{QOP}), and $N\ge m$, where $m$ is the Kraus rank of $\E$.

 \section{Orthogonality of Kraus States}

 Let 
 $$
 \E(\rho) =  (K^{(0)}\otimes I_B)\rho(K^{(0)} \otimes I_B)^\dagger  + (K^{(1)}\otimes I_B)\rho(K^{(1)} \otimes I_B)^\dagger, \rho \in \L(H),
  $$
  where $K^{(0)}$ and $K^{(1)}$ are linearly independent Kraus operators on $\L(H_A)$ satisfying (\ref{tp}).
We prove that there are Kraus operators $R^{(0)}$ and $R^{(1)}$ on $\L(H_A)$ such that (i) the quantum operator
$\E$ is also given by  $\E(\rho) =(R^{(0)}\otimes I_B)\rho(R^{(0)} \otimes I_B)^\dagger  + (R^{(1)}\otimes I_B)\rho(R^{(1)} \otimes I_B)^\dagger$, and (ii) the vectors
\begin{equation}\label{RVS}
(R^{(0)}\otimes I_B)\ket{\Psi}\quad {\rm and}\quad (R^{(1)}\otimes I_B)\ket{\Psi}
\end{equation}
are orthogonal vectors in $H$ (where $\ket{\Psi}$ is given by (\ref{GQDS}) and all of its Schmidt coefficients are nonzero).  
 
 Set $\ket{\phi_0}= (K^{(0)}\otimes I_B)\ket{\Psi} $ and $\ket{\phi_1} = (K^{(1)}\otimes I_B)\ket{\Psi}$.  Note neither $\ket{\phi_0}$ nor $\ket{\phi_1}$ is the zero vector because they are linearly independent (by Lemma~\ref{LIL}).  Assume that $\ket{\phi_0}$ and $\ket{\phi_1}$ are not orthogonal states.  (If they are, then we are done---set $R^{(0)}= K^{(0)}$ and $R^{(1)} = K^{(1)}$.)
For $\alpha = 0, 1$, define
\begin{equation}\label{Rdef}
R^{(\alpha)} = \sum_{\beta=0}^1v_{\alpha\beta} K^{(\beta)}, 
\end{equation}
where the   $v_{\alpha\beta}$'s  are complex numbers that constitute a $2\times 2$ matrix $V$.  It is easy to check  that if $V$ is unitary,  then for any operator $\rho\in \L(H)$, 
$$
\sum_{j=0}^1(R^{(j)}\otimes I_B)\rho(R^{(j)} \otimes I_B)^\dagger  = \sum_{j=0}^1(K^{(j)}\otimes I_B)\rho(K^{(j)} \otimes I_B)^\dagger.
$$
  Thus to complete our proof, we need show only that there is a unitary matrix $V$ such that the operators $R^{(0)}$ and $R^{(1)}$ defined by (\ref{Rdef}) are also such that the vectors (\ref{RVS}) are orthogonal in $H$.     

 Let $\mu$, $\nu$, and $\theta$,  denote real numbers and observe that
$$
V = \begin{bmatrix} e^{i\mu }\cos \theta & -e^{i\nu }\sin \theta \\ 
e^{-i\nu }\sin \theta & e^{-i\mu }\cos \theta\end{bmatrix}
$$
is unitary.  Thus, using definition (\ref{Rdef}), with the scalars $v_{\alpha\beta}$ determined by the preceding matrix, we have

\begin{eqnarray}
(R^{(0)}\otimes I)\ket{\Psi} &=&(e^{i\mu }\cos \theta )\ket{\phi _{0}}-(e^{i\nu }\sin \theta
)\ket{\phi _{1}}=(e^{i\mu }\cos \theta )[\ket{\phi _{0}}-(e^{i(\nu -\mu )}\tan \theta
)\ket{\phi _{1}}]\label{RO}  \\
(R^{(1)}\otimes I)\ket{\Psi} &=&(e^{-i\nu }\sin \theta )\ket{\phi _{0}}+(e^{-i\mu }\cos \theta
)\ket{\phi _{1}}=(e^{-i\mu }\cos \theta )[(e^{i(\mu -\nu )}\tan \theta )\ket{\phi
_{0}}+\ket{\phi _{1}}].\label{RT}
\end{eqnarray}
   If we can choose the scalars $\mu$, $\nu$, and $\theta$ such that  $(R^{(0)}\otimes I)\ket{\Psi}$ and $(R^{(1)}\otimes I)\ket{\Psi}$ are orthogonal in $H$, our proof is complete.

 Observe from (\ref{RO}) and (\ref{RT}) that if we desire that  $(R^{(0)}\otimes I)\ket{\Psi}$ and $(R^{(1)}\otimes I)\ket{\Psi}$ be orthogonal, then, since $\ket{\phi_0}$ and $\ket{\phi_1}$ are not  orthogonal, we must have $\cos(\theta)\ne 0$, which justifies our factorization on the right of equations (\ref{RO}) and (\ref{RT}).    Using these factored forms from (\ref{RO}) and (\ref{RT}) and writing $\mu -\nu =\xi$, we compute the inner product of   $(R^{(0)}\otimes I)\ket{\Psi}$ and $(R^{(1)}\otimes I)\ket{\Psi}$, obtaining the following necessary and sufficient condition for the inner product to be $0$ :
 \begin{equation}\label{IPR}
-(e^{2i\xi }\tan ^{2}\theta )\braket{\phi_{1}}{\phi_0}+(e^{i\xi }\tan \theta )(\braket{\phi _{0}}{\phi_{0}}-\braket{\phi _{1}}{\phi _{1}})+\braket{\phi _{0}}{\phi _{1}}=0.
\end{equation}
In the preceding equation,  the scalar products  $\braket{\phi _{\alpha }}{\phi _{\beta }}$ ($ \alpha ,\beta =0,1)$, like the $\ket{\phi _{\alpha }}$'s themselves, can be
considered known. Thus, since $\braket{\phi_1}{\phi_0}$ is nonzero, Eq.\   (\ref{IPR})  is a quadratic
equation in the complex variable $z=$ $e^{i\xi }\tan \theta ,$ so that to produce the desired orthogonal states we simply choose values of $\theta$ and $\xi$ making  $e^{i\xi }\tan \theta $ a root of this quadratic.  

\end{document}